\documentclass[a4paper,onecolumn,11pt]{article}
\pdfoutput=1
\usepackage[utf8]{inputenc}
\usepackage[english]{babel}
\usepackage[T1]{fontenc}
\usepackage{amsmath}
\usepackage{hyperref}

\usepackage{tikz}  
 
\usepackage{fullpage}
\usepackage{amsfonts,amssymb}
\usepackage{amsthm}
\usepackage{tikz}
\usetikzlibrary{calc}
\usetikzlibrary{arrows}
\usepackage{tikz-3dplot}
\usepackage{color}  
\usepackage{hyperref}
\usepackage{bm}
\usepackage{float} 
\usepackage{appendix} 
\usepackage{lscape}
\usepackage{cite}
\usepackage{mathrsfs}
\usepackage{appendix}
\usepackage{setspace}
\usepackage{mathtools}
\usepackage{authblk}

\theoremstyle{definition}
\newtheorem{definition}{Definition}[section]

\begin{document}
	
	\title{Information is Physical: Cross-Perspective Links in Relational Quantum Mechanics}
	
 \author{Emily Adlam  \thanks{The Rotman Institute of Philosophy, 1151 Richmond Street, London N6A5B7 \texttt{eadlam90@gmail.com} } \ , Carlo Rovelli \thanks{Aix-Marseille University, Universit\'e de Toulon, CPT-CNRS, F-13288 Marseille, France; Department of Philosophy and the Rotman Institute of Philosophy, 1151 Richmond Street, London N6A5B7, Canada; Perimeter Institute, 31 Caroline Street N, Waterloo ON, N2L2Y5, Canada } }

	\date{\today} 
	
	\date{\today}
	
	\maketitle

	\section{Introduction}

	Relational quantum mechanics (RQM) is an interpretation of quantum mechanics based on the idea that quantum states describe not an absolute property of a system but rather a relationship between systems.  RQM has many very appealing features. It is a realist view which is compatible with relativity; it does not require us to add anything to the existing mathematical framework of quantum mechanics;  it is a robustly naturalistic picture which does not attach any special significance to conscious minds or measurements; and it refrains from postulating unobservable, inaccessible levels of reality like hidden variables or other branches of an Everettian multiverse. Moreover, it  seems likely that RQM will still be applicable  in the context of relativistic quantum mechanics, quantum field theory and quantum gravity, whereas many other proposed interpretations of quantum mechanics face significant difficulties when we try to extend them beyond non-relativistic quantum mechanics. 
	
	However, some problems remain - in particular, there is a tension between RQM's naturalistic  emphasis on the physicality of information and the inaccessibility of certain sorts of information in current formulations of RQM. Thus in this article we propose a new postulate for RQM which ensures that all of the information possessed by a certain observer is stored in physical variables of that observer and thus accessible by measurement to other observers. The postulate of \textbf{cross-perspective links} ensures that observers can reach intersubjective agreement about quantum events which have occurred in the past, thus shoring up the status of RQM as a form of scientific realism and ensuring that empirical confirmation is possible in RQM.

Adding this postulate  requires us to clarify some features of the ontology of RQM, because it entails that not everything in RQM is relational. In this article we suggest an ontology which upholds the principle that quantum states are always relational, but which  also postulates a set of quantum events which are not relational. A quantum event arises in an interaction between two systems such that the values of some physical variables of one system become definite relative to another system, and these quantum events are observer-independent  in the sense that any other observer can in principle obtain the same information about the values of the relevant variables by an appropriate measurement on either of the systems. 

 This new postulate also provides new resources for responding to existing objections to RQM, including the solipsism objection, the preferred basis problem, and the problem of determining when a quantum event occurs. We explain  how \textbf{cross-perspective links} helps address these problems and finally we address the Frauchiger-Renner experiment in the context of RQM.

\section{RQM \label{RQM}}
According to ref \cite{1996cr}, the founding principle of RQM is the idea that \emph{`in quantum mechanics different observers may give different accounts of the same sequence of events.'} RQM has undergone significant development since this original proposal, but the basic idea remains the same: different observers may assign different quantum states to a given system and moreover in such cases all of the different assignations are equally correct, because the quantum state assigned to a system describes not only the system itself but also the relation between the system and the observer assigning the state.  There exist other interpretations of quantum mechanics which take a similar view on the relational nature of quantum states\cite{brukner2015quantum,articlebanana,demopoulos2012generalized,Janas2021-JANUQR,QBismintro} but typically these accounts regard (conscious) observers as playing some sort of privileged role. On the other hand RQM is built on strong naturalistic intuitions, and therefore in RQM the term `observer' is understood in a broad sense which allows that any physical system can be an `observer,' so we don't have to accept that consciousness plays any fundamental role. 

In this article we will take existing formulations of RQM to be characterised by the following six postulates, which are endorsed in a recent presentation of RQM in ref \cite{dibiagio2021relational}. Thus it is this specific version of RQM to which our proposed modifications apply. 

\begin{enumerate} 

\item \textbf{Relative facts:} Events, or facts, can happen relative to any physical system.

\item \textbf{No hidden variables: }unitary quantum mechanics is complete

\item \textbf{Relations are intrinsic:} the relation between any two systems A and B is independent of anything that happens outside these systems’ perspectives

\item \textbf{Relativity of comparisons:} it is meaningless to compare the accounts relative to any two  systems except by invoking a third system relative to which the comparison is made.

\item  \textbf{Measurement:} an interaction between two systems results in a correlation within the interactions between these two systems and a third one; i.e.  with respect to a third system W, the interaction between the two systems S and F is described by a unitary evolution that potentially entangles the quantum states of S and F.

\item  \textbf{Internally consistent descriptions:} In a scenario where F measures S, and W also measures S in the same basis, and W then interacts with F to ‘check the reading’ of a pointer variable (i.e. by measuring F in the appropriate ‘pointer basis’), the two values found are in agreement.

\end{enumerate}

\section{Intersubjectivity \label{intersubjectivity}}

In addition to the six postulates set out in section \ref{RQM}, a further principle that has come to be associated with RQM is the idea that information is physical. For example, Rovelli and di Biagio write, `\emph{In a naturalistic philosophy, what F “knows” regards physical variables in F. And this is accessible to W. If knowledge is physical, it is accessible by other systems via physical interactions. It is precisely for this reason that knowledge is also subjected to the constraints and the physical accidents due to quantum theory.}'
 
However, some of the types of `information' arising in RQM as formulated in section \ref{RQM} do not seem very physical. Consider for example  a  scenario in which Bob knows that his friend Alice is performing a measurement on  a system S. When Alice performs the measurement, she witnesses some measurement outcome $M_A$ and thus learns the value of some variable of the system S. But since Bob describes the whole interaction unitarily, from his point of view the interaction has only caused Alice and S to become entangled; it has not selected any one measurement result. Now suppose Bob measures S in the same basis as Alice's measurement, and hence he obtains a measurement outcome $M_B^S$ which he will interpret as providing information about the result of Alice's measurement on S.  Suppose that Bob also `measures' Alice herself and obtains a measurement outcome $M_B^A$ for the value of of some pointer variable which is supposed to be a record of her measurement result - for example, he could simply ask her what her measurement result was. So in this scenario we have three different measurement outcomes $M_A, M_B^S, M_B^A$ all supposedly providing information about the value of the same variable. What does RQM say about the relationships between these three measurement results? 

Well, clearly \textbf{Internally consistent descriptions} entails that $M_B^S$ and $M_B^A$ will agree.  But this leaves a further question about whether  $M_B^S$ and $M_B^A$ will match $M_A$. Unitary quantum mechanics does not provide any mechanism for a single measurement outcome to be selected and actualised for Alice in the first place, so it certainly cannot tell us anything about  the relationship between her outcome and Bob's outcome - this is a question which lies entirely outside the unitary part of the theory. The purpose of interpreting quantum mechanics is precisely to tell us how the unitary part of the theory relates to the measurement outcomes witnessed by observers, but RQM  as formulated in section \ref{RQM} is also silent on this question. Indeed, \textbf{Relativity of comparisons} implies that it is not even meaningful in this version of RQM to ask about the relationship between Alice's perspective and Bob's perspective, so we certainly cannot hope for any guarantee that Bob's measurement outcomes will match Alice's, since relativity of comparisons denies that there could be any fact about whether or not they match.\footnote{Refs \cite{articlebrown,2021quintet} have analysed similar cases and have similarly concluded that extant versions of RQM do not provide any reason to think that Bob's measurement outcomes will match Alice's.} We are only allowed to compare these perspectives from the point of view of a third observer, Charlie; but just as Bob is not able to find out the value of $M_A$, so Charlie is not able to find out the value or $M_A$ or the value of   $M_B^S$ and $M_B^A$, so he can't do anything to compare these values. All he can do is check whether the outcomes of his own measurements on Alice and Bob and S match, and of course \textbf{Internally consistent descriptions} entails that they always will,   but internal consistency within the results of measurements relative to Charlie says nothing about whether the  subjective experiences of Alice and Bob, as perceived by Alice and Bob themselves, are matching. So it seems that there is no way for anybody but Alice to ever find out what Alice's measurement result was. Even when Alice tries to communicate to other observers what result she saw, \textbf{Internally consistent descriptions} guarantees that everyone will always perceive her to be agreeing with them, and thus no form of communication  will ever bridge the gap between Alice's perspective and the other observers around her. Thus it seems that Alice's knowledge fails to be physical in any meaningful sense, since it does not satisfy the naturalistic criterion of Rovelli and di Biagio: `\emph{If knowledge is physical, it is accessible by other systems via physical interactions.}'

The problem is that in this scenario we are dealing with `knowledge' under two different guises - we must distinguish between the \emph{subjective} experience of  knowledge and the  way in which that knowledge   is represented in physical variables of the relevant system\footnote{We note in passing that the distinction between the subjective experience of knowledge and the physical representation of knowledge in RQM intersects in interesting ways with a number of ongoing philosophical disputes about knowledge. We do not have space in this article to do this subject justice, but we hope to return to it in future work.}. In a classical setting we can usually take it that the content of an agent's subjective knowledge coincides with the physical representation of their knowledge, because classically we usually assume that a person's subjective experiences are determined entirely by the physical state of their brain, which is regarded as an observer-independent fact.  Thus in a classical setting, if  I try to communicate something to you and my communication equipment and your perceptual equipment are working correctly, the states of our brains will become appropriately correlated, and since our subjective perspectives are assumed to be fully determined by the observer-independent physical states of our brains, it follows that our subjective perspectives will now agree and  we will have the same subjective knowledge about whatever I was trying to communicate. But postulate four of RQM as in section \ref{RQM} rules out the existence of a `view from nowhere' which would enable us to coordinate perspectives in this way, and it follows that these types of knowledge do not typically coincide in RQM: when a value becomes definite relative to Alice, she has some subjective knowledge about a measurement outcome which is not represented in  any of her physical variables which are accessible to others,   because relative to other observers she is just correlated with the system and no definite outcome has occurred.

Thus there appears to be a tension between two founding principles of RQM: the idea that information is physical and the idea that there exists no `view from nowhere' from which the perspectives of different observes can be compared.  How is this tension to be resolved? One possible response would be to argue that when Alice observes a measurement outcome her `knowledge' is somehow illusory: the real information  is in the physical correlations which are accessible to Bob and other observers, which fail to single out any one measurement outcome. But in fact Alice's subjective perspective cannot simply be disregarded in this way, because it is the subjective perspective which plays the central role in empirical confirmation. After all,    when we are trying to get empirical confirmation for a scientific theory we are necessarily doing that from within our own subjective perspective based on the world as we perceive it: even if it is the case that from Bob's point of view Alice's information  takes the form of correlations between between her and the systems that she has measured, nonetheless  that information presents itself to \emph{Alice} in the form of a description of a definite macroscopic reality in which a string of measurement outcomes has really occurred, and that is the form in which she will use it to carry out empirical confirmation. Thus in order for a scientific theory to have empirical confirmation, it must be clear about the way in which subjective perspectives are supposed to arise from the physical reality it postulates.\footnote{These concerns about empirical confirmation are discussed in greater detail in ref \cite{https://doi.org/10.48550/arxiv.2203.16278}.} 

Moreover, because postulate four prevents us from making comparisons between different subjective perspectives, RQM as formulated in section \ref{RQM} does not in general allow that a person's subjective perspective can be inferred from physical variables which are accessible to anyone else, which has the consequence that it is impossible within this version of RQM to learn about the content of anyone else's subjective perspective. As we have seen, even if Alice tries to communicate the content of her subjective perspective to Bob, she will never succeed in telling him that she disagrees with him on some measurement outcome, so he can never find out about the ways in which their perspectives differ and the ways in which they are the same. It follows that if Bob is trying to carry out empirical confirmation, he will only be able to confirm a description of his \emph{own} set of relative facts: he will never have any grounds for imagining that other perspectives are like his own in any way. Indeed, since  different versions of the same person will presumably count as different `observers,' in the RQM sense, he won't even be able to confirm that his memories match the experiences of his past selves, and thus he won't even be able to trust any relative frequencies that he may have arrived at. Thus in this version of RQM  it appears that each observer is trapped inside their own instantaneous perspective, unable to get information about what the world is like for other observers or at other times. Yet RQM is clearly intended to be a theory describing the experiences of all observers across all of spacetime, not just the instantaneous experiences of a single observer - and it's hard to see how we could be justified in believing such a theory if our epistemic circumstances are really as implied by  the six postulates above, so it would seem that RQM itself implies that we should not believe RQM!  

Thus in order for it be epistemically rational for us to believe RQM, it is necessary that there should be  some  mechanism for achieving intersubjective agreement between observers so we can have some idea of what the world is like for other observers. This indicates that the tension we have pointed out must be resolved by bringing the subjective experience of knowledge and the physical representation of knowledge back into alignment:  when an observer in RQM is involved in an interaction, the knowledge they obtain by looking at a measurement outcome must be recorded in their physical variables and must therefore be accessible to other observers. The accessibility of this knowledge will then guarantee that observers can align their perspectives by exchanging information, and therefore it will be possible to arrive at intersubjective agreement about the features of reality that we use to obtain empirical confirmation for scientific theories.

\section{Cross-perspective links}

In accordance with the discussion of section  \ref{intersubjectivity}, we suggest removing postulate four and replacing it with the following postulate. We consider that this postulate leads to a version of RQM which is more in  accordance with its underlying naturalistic motivations and its emphasis on the physicality of information. 

\begin{definition} 

\textbf{Cross-perspective links:} In a scenario where some observer Alice measures a variable V of a system S, then provided that Alice does not undergo any interactions which destroy the information about V stored in Alice's physical variables, if Bob subsequently measures the physical variable representing Alice's information about the variable V, then Bob's measurement result will match Alice's measurement result.

\end{definition} 

This postulate can be understood physically as follows. When a system Alice has information about the variable V of system S, this information is necessarily stored physically in some of the variables of Alice; and part of what it means for that information to be `physical' is that it should be accessible to other observers who have access to Alice and the ability to perform appropriate measurements. Thus as long as the information stored in Alice's variables remains intact, it follows that when Bob measures the physical variable representing Alice's information about S, then Bob should have information about the information that Alice has about S. That is to say, Bob can obtain information not only about the physical representation of Alice's knowledge but also about the content of her subjective perspective, since her subjective perspective is now understood to be encoded in physical variables which are accessible to B. 

We note that the possibility of `destruction of information' follows immediately from the two postulates employed in the original formulation of RQM in ref \cite{1996cr}: `\emph{There is a maximum amount of relevant information that can be extracted from a system,}' and `\emph{It is always possible to acquire new information about a system.}' Clearly these postulates imply that sometimes when we acquire new information about a system, some of our previous information becomes irrelevant. We emphasize that these postulates must be understood to apply individually to each observer: if Alice is in possession of the maximum amount of relevant information about S, that does not prevent Bob from obtaining some different information about S, provided that Bob does not currently have access to Alice's information about S. However, because Bob cannot have more than the maximum amount of relevant information about S, it follows that if Bob obtains some different information about S, at least some part of the information that Alice has about S becomes irrelevant to Bob, i.e. he is subsequently unable to access this information and it will play no role in determining his future interactions with S. Thus Alice's information can be `destroyed' in the sense that it becomes irrelevant to Bob - although it could potentially still be relevant to some third observer who does not have access to the information that Bob has about S, which demonstrates that the question of whether or not information has been destroyed in RQM must be relativized to an observer, as one might expect from the fact that quantum states are relativized to an observer. We can use standard quantum mechanics to determine the degree to which information is `destroyed' (relative to a given observer) in a certain interaction. Specifically, if $A_V$ is the `pointer variable' of Alice in which the outcome of her measurement on S is recorded, it follows that if Alice undergoes an interaction with Bob in which one of her variables $A_Q$, which does not commute with $A_V$, takes on a value to within precision $\delta A_Q$ , the information in $A_V$ is disturbed relative to Bob. The degree of the disturbance can be quantified by the Heisenberg disturbance relation: $\delta A_V \delta A_Q  \propto \hbar$, i.e. the disturbance to the information stored about $A_V$ is inversely proportional to the precision with which $A_Q$ has been measured.

Note that if Alice is a macroscopic system like a human being, standardly her interactions will involve position-basis variables which all commute with one another, and thus typically information stored in Alice's physical variables will be very robust. The only way to erase that information would be to exert very fine microscopic control to measure Alice in a basis other than the position basis, which is not currently within the reach of experimental technique. However, if Alice is just a qubit which has interacted with some other qubit $S$ it would be easy to measure Alice in a basis which does not commute with $A_V$ and hence destroy the information stored in Alice's physical variables about $S$, and thus information stored in the physical variables of microscopic systems is not at all robust and frequently becomes inaccessible. 

We note that the postulate of \textbf{cross-perspective links} has something in common with an earlier proposal by Van Fraassen\cite{vanFraassen2010-VANRQM}. The concerns raised by Van Fraassen in this article are similar to those which motivate \textbf{cross-perspective links}: he worries that in the earlier version of RQM, `\emph{an observer O can register a measurement outcome ... but this fact is not equivalent to O being in a particular physical state, whether relative to itself or relative to any other observer}' and thus he proposes some new postulates to answer the question `\emph{what relations are there between the descriptions that different observers give when they observe the same system?}'  For example,  he stipulates that for any systems S, O, P (witnessed by ROV), the state of S relative to O (if any) cannot at any time be orthogonal to the state of S relative to O+P (if any).   However, Van Fraassen's postulates do not fully solve the problems we have discussed here, because he uses an additional observer ROV, relative to whom these constraints hold; and if these postulates  only constrain the relationships between S, O and P relative to ROV, then they still fail to offer any grounds for a relationship between the subjective perspectives of S, O and P, and thus they still do not give observers a means of getting outside their own perspective to learn about other perspectives, which   is necessary if empirical confirmation is to be viable. Thus in our postulate of \textbf{cross-perspective links} we have refrained from relativizing the agreement between Alice and Bob to the perspective of an additional observer in the way that Van Fraassen does.

\subsection{Stable facts}

We remark that combining \textbf{cross-perspective links}  with \textbf{internally consistent descriptions} implies that if Bob measures the variable $V$ directly on $S$ instead of measuring Alice, then provided that neither S nor Alice has been disturbed since the original interaction between S and Alice,  it follows that Bob's measurement on S will have the same result as Alice's measurement on S. (Note that if S is subject to a non-zero Hamiltonian, then of course the variable $V$  should be subject to appropriate time-evolution; e.g. if Alice measures the variable $V$, and Bob's measurement of Alice takes place after a time $t$ has passed according to some appropriate clock, then Bob will need to measure a variable corresponding to $U^{-1}  V U $ rather than $V$, where $U$ is the time-evolution operator $U^{-i Ht}$). This is because  \textbf{internally consistent descriptions} implies that  if Bob measures both Alice and S the results of the measurements must match (in the notation of section \ref{intersubjectivity} we must have $M_B^S = M_B^A$) and \textbf{cross-perspective links} tells us that if he measures Alice his result will match hers (in the notation of section \ref{intersubjectivity} we must have $M_A = M_B^A$) and thus by transitivity  have $M_A = M_B^S$. So the combination of \textbf{cross-perspective links} and \textbf{internally consistent descriptions} entails that out of the substratum of relational facts we will quickly arrive at a well-established set of intersubjective facts which command agreement across many different perspectives. 
 
In particular,  \textbf{cross-perspective links} plays an important role in the emergence of a stable and \emph{shared} macroscopic reality in RQM. Ref\cite{2021sfrf} demonstrates that within RQM decoherence processes give rise to \emph{stable facts ... whose relativity can effectively be ignored.}'  These stable facts arise in the situation when Alice measures a variable V of some system S. The variable V then takes on some definite value v relative to Alice but not relative to another observer Bob who has not interacted with Alice or S. However   the value of V can be  considered stable for Bob if, in computing the probability for some other variable Q to take the value q relative to Bob during a subsequent interaction involving Bob, we can write: 

\[  P(q^B) = \sum_i P(q|v_i) P(v_i^A) \] 

 The point is that this expression looks like a classical mixture; there is no interference between branches of the superposition, and thus if this expression holds, Bob can reason as if V has some definite but unknown value relative to him. Moreover,  if Alice and Bob are macroscopic observers then generally decoherence will ensure that an expression of this kind does indeed hold  (at least approximately), and therefore most facts about variables that Alice has observed will be stable relative to Bob in the sense that Bob can treat them as classical observables. 
 
 However, it must be stressed that this  expression is a description of the situation for Bob, not for Alice - from Alice's point of view V already has a definite value at this time so there can be no nontrivial classical mixture in her description of the situation. Thus the variables $v_i^A$ appearing in this equation must be understood as facts about the result of Alice's measurement \emph{relative to Bob}, i.e. these variables do not  denote the result that Alice  has perceived herself as obtaining in the measurement that she has already performed on S.  Indeed since the value of A relative to Alice herself, as selected by the measurement she has already performed, plays no role in the expression above, it would seem that there is no connection between the `stable facts' about V relative to Bob and the  value of V that Alice herself has observed. However, once we add the postulate of \textbf{cross-perspective links}, we are entitled to replace the facts about the outcome of Alice's measurement \emph{relative to Bob} with  the facts about the outcome of Alice's measurement \emph{relative to Alice}, since the intersubjective agreement underwritten by  \textbf{cross-perspective links} assures us that there cannot be any disagreement between these sets of facts.  Thus adding \textbf{cross-perspective links} to the theory of stable facts ensures that not only does there exist a stable macroscopic reality for each individual observer, but the sets of stable facts making up macroscopic reality relative to different observers can generally be expected to agree whenever they coincide.

\section{Ontology \label{ontology}}

The ontology of RQM has been described in various different ways in the literature on the subject, but in at least some presentations it appears that the whole ontology is supposed to be relational: for example, ref \cite{2007relEPR} writes `\emph{physical reality is taken to be formed by the individual quantum events (facts) through which interacting systems affect one another ... each quantum event is only relative to the system involved in the interaction,}' and similarly ref \cite{woods} describes RQM as asserting that `\emph{there is no such thing as an absolute, observer-independent physical value, but rather only values relative to observers.'} But  with the addition of the postulate of \textbf{cross-perspective links} it no longer seems possible to insist that everything is relational - or at least, it is no longer \emph{necessary} to do so - because this postulate implies that the  information stored in Alice's physical variables about the variable V of the system S  is accessible in principle to any observer who measures her in the right basis, so at least at an emergent level this information about V is an observer-independent fact. This suggests  that the set of `quantum events' should be regarded as absolute, observer-independent features of reality in RQM, although quantum states remain purely relational. Thus we continue to endorse the sparse-flash ontology for RQM as advocated in refs \cite{blueskies,sep-qm-relational}: however we now regard the pointlike quantum events or `flashes' as absolute, observer-independent facts about reality,  rather than relativizing them to an observer.
 
As in previous versions of RQM, a quantum event is understood to arise in an interaction between two systems in which the variables of one  system take on definite values relative to the other, and vice versa. For example,  suppose there is an interaction between Alice and a system $S$ which takes the form of a `measurement' of a variable $V$ of the system $S$: then the corresponding quantum event can be loosely characterised as `variable $V$ taking value $v$ relative to Alice,' where the probability for the  value  $v$ is given by the Born rule in the usual way. We reinforce that the phrase `relative to Alice' here does not indicate that the event itself is relativized to Alice; the \emph{event} is an absolute, observer-independent fact, but the \emph{value} $v$ is relativized to Alice because at this stage Alice is the only observer who has this information about $S$, although other observers could later come to have the same information by interacting appropriately with either Alice or $S$. 
 
 Evidently RQM leans strongly on the notion of a `system' and as noted in ref \cite{woods}, it may seem hard to understand what a system really is if everything is supposed to be relative to a system! However, postulating a set of quantum events which are not relative to anything helps address this question: now a `system' can simply be identified with a set of quantum events which are related to one another in certain lawlike ways, as captured by the formalism of quantum mechanics. Each system is characterised by   an algebra of physical variables, i.e. the set of variables which can take on values in quantum events associated with the system.\footnote{In a Humean approach to RQM it might be tempting to insist that every variable of the system must take on a value at least once in the set of events associated with a system. However, for non-Humeans there is no reason to make this stipulation, since  the algebra associated with the system may be understood as a modal fact encapsulated in the quantum mechanical laws describing the relations between events, so it does not have to be simply read off the set of events associated with the system.} Recall that every  quantum event involves two systems interacting, and therefore different systems do not have to be associated with   disjoint sets of events, although no two systems will be associated with the \emph{same} set of events. We note that there may appear to be some circularity here: previously events have been defined as interactions between systems, and now systems have been defined as sets of events. However, there is not actually any circularity if we start from the idea that  RQM is a theory of a set of quantum events related  to one another in lawlike ways: it then turns out that the lawlike relations work in such a way that it is possible to define `systems' such that  that every event can be regarded as an interaction between two systems, so the notion of a system is not necessarily fundamental, but rather is used as an interpretative tool to help us  make sense of the set of quantum events.

We reinforce that in this picture,  the variables of a system take on values only during a quantum event, and at all other times they have no values. This is the way in which RQM makes sense of the Kochen-Specker theorem\cite{KochenSpecker} and other contextuality theorems\cite{Spekkens}. The variables of a system therefore do not constitute a state, since they do not continue to have a definite value for any finite interval of time. If some variable $V$ of a system $S$ takes on a definite value in the course of an interaction with another system, then if another system subsequently interacts with $S$ in a way that can be regarded as a `measurement' of the same variable, it follows from \textbf{cross-perspective links} that  $V$ will take on the same definite value in the second interaction. However, RQM tells us that this information about the value of $V$ does not have to be carried from one interaction to another by a mediating physical state: rather we can think of each interaction involving $S$ as `looking back' at the most recent interactions involving $S$ to determine the outcome of the new interaction. In this sense, systems do not have states, they just have histories.  

We note that the postulates of RQM as set out in ref \cite{1996cr} can be used to determine which recent interactions must be taken into account, for the two postulates together entail that `\emph{when new information is acquired, part of the old relevant-information becomes irrelevant}.' In the language we have used here, this means that when a new interaction occurs, one or more earlier interactions cease to matter, in the sense that future interactions will no longer depend on them. So each interaction involving $S$ only has to `look back' at a finite number of recent interactions involving $S$: most of $S$'s history will be irrelevant to the outcome of the new interaction.

Although strictly speaking the variables of a system do not have values in between interactions, nonetheless one may sometimes wish to speak colloquially of variables `having values' at other times. In this colloquial sense, in the case where the Hamiltonian of the relevant system is zero so there is no time-evolution, what it is  for a  variable of $S$ to `have a value' is simply for that value to be realised in the most recent interaction involving $S$:  for example, $S$ `has a value' for the quantity $V$ in the colloquial sense if the most recent quantum event for that system was an interaction in which the quantity $V$ took on some definite value.   If there is a non-zero Hamiltonian,  $S$ `has a value' for the quantity $V$ in the colloquial sense if the most recent quantum event for that system was an interaction in which the quantity $U  V U^{-1} $ took on some definite value, where $U$ is the time-evolution operator $U^{-i Ht}$ and $t$ is the time since the interaction, relative to some appropriate choice of clock (note that we apply the \emph{inverse} time-evolution operator to $V$ in order to obtain the relevant quantity at the time of the interaction). This locution makes sense precisely because \textbf{cross-perspective links} ensures that subsequent measurements of the same quantity will reveal the same value.   The interaction in which $V$ takes a definite value supersedes any previous interactions in which a variable that does not commute with $V$ took on a value, so the previous information is `destroyed’ in the sense that it can no longer be accessed by future measurements on $S$. Thus at any given time, most of the variables of $S$ will not `have a value' even in this colloquial sense. 

It should be noted this ontology for RQM makes a departure from some previous approaches to RQM which have tended to associate it with a fairly radical form of metaphysical indeterminacy. For example, ref \cite{CALOSI2020158} distinguishes between `gappy metaphysical indeterminacy' (where no determinate of a determinable is instantiated) and `glutty metaphysical indeterminacy' (where more than one determinate of a determinable is instantiated) and argues that both of these occur in RQM. By contrast, the version of RQM we have presented here exhibits gappy metaphysical indeterminacy, since variables have no definite values in regions between quantum events, but \emph{not} glutty metaphysical indeterminacy, because \textbf{cross-perspective links} ensures that whenever two observers both know something about the value taken by a variable in a given interaction, their knowledge will always match, so  we will never have a case where a physical variable takes two different values relative to different observers (either the variable takes the same  value relative to both observers, or it takes no value at all relative to one of the observers, if that observer  currently does not have any  information about it). Of course it is still the case in our version of RQM that a given system may be assigned two different quantum states by different observers, but not because of some kind of indeterminacy - quantum states differ between different observers simply because a quantum state describes the \emph{relation} between the observer and the system rather than an absolute feature of the system. In a similar way, the person who I refer to as `my mother' is probably not the same as the person who you refer to as `my mother,' but this is does not mean that there is any metaphysical indeterminacy about the use of the term `my mother' - the point is that the term `my mother' describes not an absolute feature of a person but a \emph{relation} between the speaker and the person described, and my relation with that person is not the same as your relation with that person. 

We also note that unitary quantum mechanics provides us with what might be described as a `patchwork' account of the distribution of quantum events, with each individual relational description characterising the relation between some particular event and the most relevant events in the past, but nothing in the theory characterising the distribution as a whole. It then seems natural to wonder if it might be    possible to give a more unified description - something like a probability distribution which assigns probabilities to  the distribution of quantum events across the whole  universe, from which the individual relational descriptions could be extracted. However as noted in ref \cite{blueskies}, RQM indicates that this unified description certainly cannot be derived from `the quantum state of the full universe,' because quantum states are by definition relational and there is nothing for the quantum state of the whole universe to be relativized to. So perhaps we should conclude that there is actually no unified description of the full set of quantum events and the `patchwork' description is truly fundamental. Or perhaps some way of giving a unified description will emerge from ongoing research on quantum gravity and quantum cosmology - for example, ref \cite{2019switch} suggests an interpretation of the wave function of the universe in the framework of Dirac quantisation where it is understood as `\emph{a perspective-neutral global state, without immediate physical interpretation, that, however, encodes all the descriptions of the universe relative to all possible choices of reference system at once}.' This approach might be seen as consistent with RQM provided we are clear that the universal wave function obtained during Dirac quantisation is not a quantum state in the ordinary sense, since it is not relativized to anything. But in any case, RQM as it currently stands already provides us with a coherent understanding of standard quantum mechanics as  a means of locally navigating the set of quantum events: we can continue to assert that `\emph{Quantum mechanics provides a complete and self-consistent scheme of description of the physical world, appropriate to our present level of experimental observations}'\cite{1996cr} because this unified description, if it exists, would certainly go beyond our present level of experimental observations.

\subsection{Relational Quantum States}

The slogan that `systems do not have a state, they just have a history' provides a straightforward way of understanding the relational nature of quantum states: we assign a quantum state based on what we know about recent history, so if you and I have different information about the recent history of a system, we will assign different states. It also helps us see why wavefunctions should be updated after measurements in the context of RQM, even though there is never any physical collapse or breakdown of unitarity. For in this picture,  the purpose of the quantum state assigned by Alice to $C$ is to describe the information that Alice has about the history of $C$. Thus when a new quantum  event involving both Alice and $C$ occurs, it must be added to the information that Alice has about the history of $C$, and so of course Alice has to start again using a different quantum state which takes account of the most recent event. Thus there must be a state update but that  update is not a   physical process located in spacetime.

However, this way of thinking about relational quantum states might give the impression that they are purely epistemic, i.e. they merely describe the information that different observers have about history. And this may seem to give rise to a puzzle. If Bob does not know the result of Alice's measurement of the variable $V$ on the system $S$, he describes Alice and $S$ as being in a superposition of all the different possible values of the variable that she measured. But we know that in the interaction of Alice and $S$ a single value of $V$ has become definite relative to Alice, and this value is an observer-independent physical fact in the sense that if Bob were to measure Alice or $S$ in the same basis he would obtain a result that agreed with Alice's result. But nonetheless, if Bob does not perform this measurement and instead chooses to perform interference experiments on Alice, he is able to see interference effects. Typically we imagine that interference occurs because the superposition is physically real and it represents the fact that all of the possible values of $V$ in some sense coexist, thus explaining why they are able to `interfere' with one another. So if in fact one particular value of $V$ has been actualised and the superposition state is merely an expression of Bob's lack of knowledge about this value of $V$,  how is it that interference effects can occur? 

The answer lies in the principle that all information is physical. The quantum state assigned by Bob is indeed an expression of his knowledge about the recent history of $S$, but that knowledge is not an abstract disembodied idea: what he does and does not know about the recent history of $S$ is a result of his own  interactions with $S$, and thus his knowledge is stored in his physical variables in just the same way as the outcome of Alice's measurement on $S$ is stored in hers. Thus the relational quantum state assigned by Bob to $S$ is not just a description of what Bob knows about the history of $S$; it characterizes the \emph{joint} history of Bob and $S$, i.e.  all their direct and indirect interactions in the past (where by `indirect' interactions we mean cases where Bob obtains information about $S$ not by interacting with $S$ directly but by interacting with other systems which are connected to $S$ by some continuous chain of interactions; \textbf{cross-perspective links} ensures that possession of information about $S$ is transitive so indirect interactions of this kind are possible). In this way, Bob's quantum state does indeed describe his knowledge, but that knowledge is a feature of physical reality so the quantum state is also a feature of physical reality.

Thus the mantra that `information is physical,' offers a new perspective on the traditional dichotomy between `epistemic' and `ontic' views of the quantum state\cite{Spekkensepistemic,Quanta22} - `epistemic' approaches being those which say that the quantum state is merely a description of knowledge, and `ontic' approaches being those which say that the quantum state is an element of reality. For if we accept that information is always physical, then \emph{knowledge} is physical, and therefore there can be no sharp distinction between epistemic and ontic approaches; if the quantum state is a description of knowledge, it is a description of something physical and therefore it is ontic. Of course, traditionally ontic approaches  insist that the quantum state is an ontic state \emph{of the quantum system}, not of the observer who assigns the quantum state, but naturally proponents of a relational view will reject that distinction: RQM tells us that in order to understand the nature of the quantum state we must consider the observer and quantum system together, and then we will appreciate that the quantum state is ontic in the sense that it describes the information stored in the variables of the observer and the way in which that information shapes the possibilities of future interactions between the observer and quantum system. Thus in the relational picture there is no need to choose between epistemic and ontic views of the quantum state - both are correct! 

This observation leads us to a somewhat different story about the nature of quantum interference. Interference does not occur in the scenario described above because all of the possible values of $V$ coexist; instead interference  is to be understood as   \emph{destroying} the information about which one of the possible values for $V$ has been realised in Alice's interaction with $S$.  When Bob performs an interference experiment on Alice and $S$, this interaction will give rise to a quantum event which in which some variable of the joint system of Alice and $S$ takes on a definite value relative to Bob, and if Bob chooses a basis which is orthogonal to the pointer variable which would reveal the outcome of the measurement of $V$ to him, then this new quantum event supersedes the previous event in which $V$ took on a value relative to Alice. Then assuming that no other system has interacted with Alice or $S$ since their last interaction,  there is subsequently no way for any system to get information about the value of $V$ that was realised in the interaction, so to all effects and purposes it is as if no definite value  was realised at all. 

Thus the reason that interference can still occur even though $V$ has taken on a definite value is  a reflection of the fact that in RQM, an interaction between two systems cannot be regarded as simply probing the state of one system, or indeed probing the state of both systems: instead the interaction depends directly on the recent history of direct or indirect interactions between these two systems. This is simply  a fact about the distribution of quantum events in RQM. Any scientific theory is ultimately designed to predict some distribution of events, i.e. measurement outcomes, and therefore any scientific theory must propose some model for the way in which events depend on one another:  the model in which events come to depend on one another based on information carried through time  by mediating physical states is one possible approach, but it is not the only conceivable way that events could depend on one another and indeed it is by now clear that this model does not work very well for quantum mechanics. RQM offers another type of model in which events depend on past events in ways that are more complex than the simple `state' model would lead us to expect.

\subsection{Time symmetry \label{time}} 

In section \ref{ontology} we suggested that `\emph{we can think of each interaction involving $S$ as `looking back' at the most recent interactions involving $S$ to determine the outcome of the new interaction}' and that `\emph{systems do not have states, they just have histories.}' This description is distinctly asymmetric in time, which may seem surprising. For the mathematics of quantum mechanics is perfectly time-symmetric (it has recently been shown that this is true even if we include the Born rule\cite{2021aot}) and therefore one might naturally think that an interpretation of quantum mechanics should exhibit the same symmetry, otherwise we will have questions to answer about why the underlying ontological time-asymmetry does not show up in any of the mathematics. 

However, the ontology we have suggested need not actually be understood in a time-asymmetric way. For the requirement of \textbf{cross-perspective links} can also be written in the following form:   `if $S$ is involved in an interaction with $A$ in which the variable $V$ takes on some definite value relative to $A$, and then the next event involving $S$ or $A$ is also an interaction in which the variable $V$ takes on some definite value, these two values match.' This is a perfectly time-symmetric requirement: it is not the case that the earlier definite value causes the later definite value or vice versa, it's simply the case that the definite values are required to be the same. So there is no need to imagine that the set of quantum events being generated in some particular temporal order: they can simply be generated `all-at-once' in a time-symmetric and atemporal fashion. 

Similarly, although we have argued that the quantum state assigned by $A$ to $S$ should be regarded as a descriptions of the history of interactions between $A$ and $S$, that does not mean there is anything special about past interactions as compared to future interactions. The reason quantum states describe histories and not futures is simply that the theory of quantum mechanics has been developed under circumstances in which we typically know about the past and not the future, and thus we have arrived at a theory which predicts future events on the basis of past ones. However, if we typically knew about both past events and also future events then we could have arrived at a different theory which predicts the definite value that will be taken on in a given interaction between $A$ and $S$ based on both their past interactions and also their future interactions. Indeed, there exists a formalism for exactly this purpose - the two-state vector formalism\cite{Aharonov} defines both a forwards-evolving state and a backwards-evolving state and then uses the ABL rule\cite{Mohrhoff_2001} to produce predictions for the outcome of a measurement conditional on both the past and the future.  In relational terms, this means that if   we knew both the history of interactions of two systems up to some point and also all their future interactions after some point, we would be able to write down two relational quantum states which would describe the way in which the interaction `looks back' at the recent history and also `looks forward' at the immediate future to determine the value which becomes definite during the interaction. So in fact, the correct thing to say is that in an interaction the definite value taken on in the interaction depends on the whole history of interactions between the systems, both past and future; it is just that we are not typically in a position to write down the relational state encoding the future interactions. 

\subsection{Locality} 

The version of RQM that we have set out here is nonlocal in a straightforward way. For we have emphasized that in RQM information does not need to be carried from one interaction to another by a mediating physical state: instead an interaction `looks back' at recent history and thus the outcome of an interaction depends \emph{directly} on past quantum events. Indeed, this feature follows directly from the founding principle of RQM, i.e. the idea that `\emph{In quantum mechanics different observers may give different accounts of the same sequence of events.}' In particular, we have emphasized that although in the  version of RQM proposed here it is possible to arrive at intersubjective agreement about quantum \emph{events}, different observers will in general have different stories to tell about quantum states and thus about the temporal development of a system in terms of the evolution of quantum states. Moreover, RQM also insists that no single observer is right - each account is a correct characterisation of the relationship between the observer in question and the system to which the state is being assigned. Thus, assuming that the observers all live in the same world (i.e. we are not dealing with some kind of Everettian arrangement),  we cannot say that the observers are all describing a unique  process which literally takes place in spacetime, since their accounts would be mutually contradictory. This suggests that in RQM temporal evolution need not be taken literally: rather we can suppose that quantum events depend directly on one another, and the evolution processes described by different observers simply capture the inferences that these observers can make about how their past interactions determine their possible future interactions with a given system.

Moreover, as noted in section \ref{time}, we do not need to think of the set of events as being generated in some particular temporal order. In fact, we can say something even stronger: if we want to maintain relativistic covariance then we \emph{cannot} think of the set of events as being generated in some particular temporal order. This point has been noted in the context of other ontologies consisting of pointlike events - for example, Esfeld and Gisin note that the Bell flash ontology is relativistically covariant only if `\emph{one limits oneself to considering whole possible histories or distributions of flashes in spacetime, and one renounces an account of the temporal development of the actual distribution of the flashes in space-time.}'\cite{Gisin2013} Thus it seems that RQM is most compatible with a metaphysical picture in which where the laws of nature apply atemporally to the whole of history, fixing the entire distribution of quantum events all at once. And as argued in ref \cite{adlam2022roads}, a metaphysics in which the laws apply in this way is automatically nonlocal:   if we accept that the laws assign probabilities  directly to the entire distribution of events across the spacetime, there is no need for information about one part of spacetime to be carried to other parts by a physical beable propagating through spacetime. After all, in this picture the outcome of  a given quantum event is not determined by the information locally available at that event, but rather by the probability distribution assigned from the outside, which necessarily contains information about all the other quantum events happening elsewhere and at other times. 

However, we reinforce that this kind of nonlocality does not involve superluminal signalling or some kind of collapse which takes place on a spacelike hyperplane: because RQM does not take temporal evolution literally, it is not required to tell any story about the spatiotemporal unfolding of beables in between quantum events, or to say anything about the path along which an influence travels from one quantum event to another.  Thus although there is nonlocality in RQM, it is not of the objectionable kind that involves hidden influences or preferred reference frames, and thus there is no particular reason to try to avoid this sort of nonlocality.

\section{Some Objections} 

In this section, we describe some objections that have been made to the version of RQM set out in section \ref{RQM}, and we show that the postulate of \textbf{cross-perspective links} helps address some of these objections.

\subsection{Solipsism} 

RQM has been accused of engendering a form of solipsism: for example, Pienaar suggests that it gives rise to an ontology of `\emph{island universes.}'\cite{2021quintet} This criticism had some bite in the context of previous versions of RQM where it seemed that there was no mechanism by which observers could get information about the contents of other observers' perspectives. However, this is no longer the case once we explicitly add the postulate of \textbf{cross-perspective links}, since observers can now achieve intersubjective agreement about most quantum events by means of appropriate measurements on one another. Thus this version of RQM is no more solipsistic than ordinary classical physics - in RQM observers can find out about others' perspectives by means of direct communication and thus they can share information about their experiences of reality for the purpose of doing science.

\subsection{Systems and subsystems \label{systems}}

\textbf{Cross-perspective links} also plays an important role in clarifying some possible confusions around the relationship between systems and their subsystems. For note that every particle in my brain counts as an `observer' according to RQM's broad definition of that term, and moreover    in the version of RQM set out in section \ref{RQM}, since we are not allowed to make any direct comparisons between the sets of relative facts associated with    different `observers,' we cannot expect that the facts relative to all of the particles in my brain will ever agree on the result of any given measurement. However, when I perform a measurement, I always observe a single definite result; so if the particles in my brain do not in general agree on the result of a given measurement, it follows that my perspective cannot simply be understood as arising out of the collection of the perspectives of  all  the particles in my brain, since this would not give rise to a definite perspective at all. So on this construal of RQM it is somewhat mysterious to understand the nature of my perspective and what physical grounding it could possibly have. 

On the other hand, if we include the postulate of \textbf{cross-perspective links}, then when the particles in my brain interact they can be expected to exchange information such that their sets of relative facts become aligned. Moreover, decoherence ensures that information will be exchanged on a very short time-scale, so information possessed by one of the particles in my brain  will very quickly be possessed by all of them (at least in the dynamically favoured decoherence basis). Therefore in this version of RQM it is now feasible to suppose that  the perspective of a conscious observer simply emerges from the collection of the perspectives of all the particles in their brain - roughly speaking, a variable $V$ of a system $S$ will have a definite value $v$ relative to me if variable $V$ has the definite value $v$ relative to most of the particles in my brain (or perhaps just in some particularly relevant section of my brain - we would have to turn to neuroscience to determine how much of the brain should be included).

\subsection{Which variable?}

One objection that has been made to previous versions of RQM is that in general an `interaction' will not have the form of a measurement, and therefore it will not single out uniquely a variable of one system which should take a definite value relative to the other system during the interaction. In particular, refs \cite{mucino2021assessing, https://doi.org/10.48550/arxiv.2107.03513} note that  we can always rewrite an interaction Hamiltonian in a different basis, and a Hamiltonian which looks like it describes a measurement of variable $V$ in one basis will typically look like it describes a measurement of some other variable $V'$ when we write it in a different basis. So pure unitary quantum mechanics does not  suffice to determine which variable in particular should take on definite values during an interaction which leads to a quantum event.  

We see two options for RQM to respond to this objection. The first is to stipulate a preferred basis and insist that this is always the basis which takes on definite values during an interaction. For example, it has been noted in the context of the de Broglie-Bohm interpretation that ultimately all measurements are measurements of position, and therefore for the purpose of explaining our definite macroscopic experiences it is enough to ensure that some beables have definite values of position at least during measurements. Thus in principle one could imagine a version of RQM in which systems always take in a definite value relative to one another in the \emph{position basis} during an interaction, and no variables ever become definite in any other basis. However, we do not find this solution appealing, because the preferred basis does not seem to be directly revealed by anything in the empirical data, so it would in a sense be an unobservable, inaccessible feature of reality, and RQM aims to reject such features. In addition, we think it is  clear that decoherence must play some role in the emergence of a definite macroscopic reality, so we  do not consider it reasonable to expect that unique definite values will arise in fundamental interactions before any decoherence has taken place. 

The alternative is to agree that   quantum events do not typically have the simple form `variable $V$ taking value $v$ relative to Alice.' Rather they must have a conjunctive form:  `variable $V_1$ taking value $v_1$ relative to Alice, and variable $V_2$ taking value $v_2$ relative to Alice,  ....' and so on, specifying definite values for each of the variables singled out by the interaction Hamiltonian in all of the different possible bases for it. The probability distribution over definite values in each disjunct would again be given by the Born rule, and the values in each conjunct would be probabilistically independent. Now, this solution might seem to undermine the claim that RQM can explain why measurements have definite outcomes. However, RQM need not insist that an interaction singles out a unique value when the two systems involved are, for example, qubits. After all, standard quantum mechanics does not say anything about  what happens when one qubit `measures' another qubit, so there are no predictions here to reproduce. At this point, it is important to recall the distinction we made earlier between the subjective experience of knowledge and the physical representation of knowledge as information stored in the values of the variables of a system. As noted by Rovelli and di Biagio in ref \cite{dibiagio2021relational}, a qubit does not have a subjective experience of knowledge, it merely has information stored in the values of its variables. Indeed, as shown in ref \cite{https://doi.org/10.48550/arxiv.2107.03513}, it's not possible that qubits  have knowledge in the ordinary sense of the word: we could allow that they have knowledge `\emph{only at the expanse of a radical revision of the notion of `knowledge about a physical system'},' which would evidently be incompatible with the idea that they have subjective experiences of knowledge like our own. Thus the information that a qubit has about another qubit need not be such that we can understand what subjective experience of knowledge it would correspond to, and therefore RQM is under no obligation to solve the preferred basis problem in the case of interactions between individual fundamental particles. RQM need only show  that  in the limit as one of the systems involved becomes macroscopic, then there is a unique choice of variable which takes definite values in the interaction, in order that macroscopic conscious beings like ourselves can have definite experiences.

It is clear that decoherence should play some role in this story. And in fact, decoherence provides exactly what it needed here: it picks out a basis which is dynamically favoured and then disseminates information   stored in that basis through the environment. Recall that we saw in section \ref{systems}  that the perspective of a conscious observer must be regarded as an emergent phenomenon which arises from the combination of the perspectives of each of its constituent subsystem; thus the contents of the observer's perspective are defined not by the information associated with a single fundamental particle, but by the information that has been disseminated through their brain by decoherence processes. Typically we would expect that the decoherence basis will favour at most one of the variables that took on a value during the  original interaction, so decoherence effectively selects one variable out of the conjunction of variables that appeared in the original quantum event, and it is that variable which then has a definite value in the perspective of the conscious observer. Of course the decoherence process is not perfectly well-defined - there is no exact line between `decohered' and `non-decohered' - but that is not a problem because consciousness also does not seem to be perfectly well-defined: to our best current understanding it appears to be  some kind of emergent high-level feature of reality, so we are certainly entitled  to suppose that consciousness can emerge only when enough decoherence has occurred to single out a well-defined preferred basis.

In more detail: consider a Stern-Gerlach experiment such that at the end of the experiment, the atom involved interacts with particles on the screen. We acknowledge that the interaction Hamiltonian will not in general single out a unique variable of the atom, and therefore in this interaction some set of variables of the atom   take on definite values relative to the particles in the screen, with probabilities for each variable given separately by the Born rule. Now the particles on the screen interact with photons which in turn interact with receptors in my eyes and thus information spreads via decoherence processes through the particles of my brain. In RQM terms, this involves a large number of quantum events in which the particles in my brain undergo interactions and thus share some of the information stored in their physical variables. It has been established that at least in the case of non-relativistic quantum mechanics, the dynamical processes involved in decoherence primarily favour the dissemination of information in a coarse-graining of the position basis\cite{2012deco}, and therefore  eventually a significant number of particles in my brain will share the same information about the definite value of the atom \emph{in the coarse-grained position basis} - i.e. the information about where the atom ended up on the screen. The information about the definite values in all the other bases that were realised in the original interaction does not get disseminated in the same way because these bases are not dynamically favoured by the relevant decoherence processes. Indeed, because decoherence plays the role of a `measurement' of the definite values in the position basis, the information in the other bases necessarily becomes inaccessible, so no future interactions can obtain information about the definite values that were realised in all the other bases. Thus assuming that my conscious experience emerges from the unified perspectives of the particles in my brain, the definite value that I will become aware of is the one on which a significant number of particles in my brain agree - so I will have the experience of seeing a point in a particular coarse-grained position on the detector screen. My subjective experience of knowledge of this position is encoded in the physical variables of the particles in my brain, i.e. the values that their variables took on in their most recent interactions. 
 
This is of course somewhat similar to the account given by the Everett interpretation\cite{Wallace}, which likewise contends that the perspectives of observers emerge through the process of decoherence. But in the Everett interpretation, there is no mechanism for a definite value to ever be selected in any basis,  so what decoherence yields is a set of (approximately) distinct observers, one for each diagonal element in the decohered density matrix, each of whom has a well-defined perspective containing a different value of the variable being measured. Whereas in RQM, during an interaction every individual variable takes on a well-defined value relative to the measuring system, and  therefore decoherence will yield \emph{one} observer corresponding to exactly one of the diagonal elements in the decohered density matrix, whose perspective contains that one unique value of the relevant variable. Thus we reinforce that in RQM the role of decoherence is not to explain the occurrence of definite values or the breakdown of unitarity, which of course decoherence alone cannot do: the actualisation of definite values occurs in RQM without decoherence, but is it decoherence which selects  \emph{one particular} definite value out of the various values actualised in the interaction, such that macroscopic observers will observe a measurement result in a unique basis. 

 \subsection{When do events take place?}
 
A similar objection involves the concern that in general the time of a quantum event will not be well-defined. For example, ref \cite{mucino2021assessing} objects that `\emph{value acquisition occurs when interactions happen. It seems, then, that the framework needs for there to be a well-defined moment at which each interaction takes place; otherwise, the proposal becomes vague and loses all strength.}' The idea here seems to be that it is a defining property of an event that it should have a definite, well-defined spacetime location. But this is an intuition which arises from the assumption that we are working with a well-defined background spacetime - there is an alternative view which suggests that in fact, spacetime should be understood to emerge from a background of quantum events. This idea has a long history\cite{Capellmann2021-CAPSIQ}: it was present in the writing of many of the founders of quantum theory, particularly Max Born, who believed that `\emph{One should not transfer the concept of space-time as a four-dimensional continuum from the macroscopic world of common experience to the atomistic world; manifestly the latter requires a different type of manifold}'\cite{Bornletter}. The same idea has reappeared in several modern approaches to quantum gravity which propose a discretized spacetime\cite{cc,Sorkingeometry}. Thus since   RQM is ultimately intended to function not as an interpretation of non-relativistic QM on a fixed background, but rather as an interpretation which will work for quantum gravity, we should not necessarily expect to have a well-defined  background spacetime on which the events postulated by RQM are defined: rather, spacetime itself can arise from these events, as in ref \cite{vidotto2013atomism} - roughly speaking, the idea is that the notion of a `physical system' is replaced with a `spacetime region' and the notion of `interaction' is replaced with `adjacency,' so \emph{`variables actualize at three-dimensional boundaries with respect to (arbitrary) space-time partitions.'} \cite{blueskies}  Thus RQM does not need to insist that quantum events occur at well-defined spacetime locations: all that is necessary is that a well-defined spacetime should appear in some emergent limit. 

Furthermore, we reinforce that  the mantra `information is physical' applies to all kinds of information, including information about the time of an event. Thus assigning a time to an event has no meaning in this setting unless that time is recorded in the physical variables of some kind of clock. Since an interaction by definition involves only the two systems concerned, it follows that the time of an interaction between two systems must be defined with respect to the reading on an internal clock associated with one of the systems. Moreover, it seems natural to expect that the reading on this clock should be treated like other physical variables - it takes on a definite value during an interaction and has no value at other times. Thus in general a quantum event might actually take a form which looks something like `variable $V$ taking value $v$ relative to Alice when Alice's clock observable is equal to $t$' - corresponding, roughly, to Alice measuring the variable $V$ of system $S$ and the time variable $T$ of her clock and getting the joint outcome $(v, t)$. So after this event, only systems $S$ and Alice have information about the time of the event; other systems can only know the time of the event insofar as they can infer it from their own previous interactions with $S$ and/or Alice. However, since the information about the time is stored in the physical variables of $S$ and Alice, other observers can measure the relevant pointer variable and thus, because of \textbf{cross-perspective links}, observers can arrive at intersubjective agreement about the times at which past events occurred. 

We note that there are indications from standard non-relativistic quantum mechanics that this is the right way to think about the time of events. For example, consider the technique of `time-bin-encoding,'\cite{PhysRevLett.93.180502,PhysRevLett.111.150501} where a photon is prepared in a superposition of states $| 0 \rangle$ and $|1 \rangle$, where $|0 \rangle$ corresponds to the photon exiting the interferometer at a certain time and $| 1 \rangle$ corresponds to the photon exiting the interferometer at a later time. Here, the time of the event is used in exactly the same way as variables like position, spin, charge, energy level and so on in other methods of preparing qubits. That is to say, ignorance about the time of an event can be used to create a superposition state in just the same way as ignorance about those other sorts of variables, and the resulting state will exhibit quantum phenomena like   interference in the usual way. This underscores the fact that the time of an event is a physical variable which must be treated like all other physical variables in RQM.

There  exist various formalisms aiming to define time in a relational manner which could be employed here to further characterise the time at which a quantum event occurs.  For example, the Page-Wootters formalism\cite{2021gpr,PhysRevD.27.2885,2020cqtd}  is an approach to defining relational time in the context of Dirac quantisation. In this approach we choose a subsystem $C$ of the universe to act as a clock and then define a `time observable' $E_C(t) = |t \rangle \langle t |$ which acts on the clock system $C$ and which transforms covariantly with respect to the group generated by the Hamiltonian $H_C$ of that subsystem, i.e. $E(t + t') = e^{-iH_C t'} E(t) (e^{-iH_C t'})^{\dagger}$. Using this time observable and the universal quantum state $|| \Psi \rangle$ (which may in the context of RQM be regarded as just a mathematical tool with no physical meaning) we can define a state for the rest of the world,  $W$, conditional on $C$ reading time $t$: $| \psi_W(t) \rangle = \langle t | \otimes I_W || \Psi \rangle$. In the RQM formalism, we may imagine that the clock $C$ is associated with our observer Alice, and we may suppose that Alice interacts with some other subsystem of the universe $S$; then if Alice's interaction with $S$ takes the form of a measurement of some variable $V$, the probability that she sees the value corresponding to $| v \rangle \langle v |$ conditional on her clock reading $t$ is $\langle v |_S Tr_{W - S} ( | \psi_W(t) \rangle )$ where the trace is taken over the whole `rest of the world' minus $S$. Thus this framework allows us to extend the relational probability distributions for measurement outcomes defined by the Born rule to joint probability distributions over measurement outcomes and the time reading to which those measurements outcomes will correspond.

\section{The Frauchiger-Renner Experiment} 

The Frauchiger-Renner experiment\cite{2018qtcc} is a thought experiment which provides a particularly clear illustration of the ambiguities inherent in textbook unitary quantum mechanics. It describes  two experimenters (Alice and Bob) performing measurements inside boxes and two other experimenters (Wigner and Xena) performing measurements on the experimenter inside the boxes. We will not reproduce the details of the argument here: the key point is that at the end, Wigner performs a certain measurement $M$, and if all of the observers analyse the experiments and their observations using unitary quantum mechanics, it turns out that there are some circumstances in which Wigner can be sure that Bob, inside his box, has predicted that Wigner will definitely obtain the outcome $0$ to measurement $M$, but also Wigner himself must predict that it is still possible for him to get outcome $1$ to measurement $M$. That is, two agents, both applying quantum mechanics correctly, come up with incompatible  predictions for the outcome of the same measurement.  

It is not difficult to understand where this apparent paradox comes from. Bob models himself as an observer and thus when he performs a measurement inside the box, he adds something like a wavefunction collapse into his quantum description corresponding to the measurement outcome that he has seen. Wigner on the other hand models Bob as a quantum system, and thus represents Bob's measurement as Bob becoming entangled with the measured system with no wavefunction collapse, so Bob remains in a coherent superposition until the time of Wigner's measurement. Thus at the time of Wigner's measurement $M$, Wigner and Bob  assign different quantum states to the overall system and thus they make different predictions for the outcome of the measurement. 

Now, if we think that measurement $M$ must have a  definite outcome, then clearly we must conclude that either Wigner or Bob is wrong. Wigner is wrong according to any interpretation which tells us that people cannot be in superpositions - e.g. gravitational collapse interpretations and spontaneous collapse interpretations - because those interpretations affirm that there will be an actual or effective collapse of the wavefunction at around the time of Bob's measurement so that Bob will no longer be in a coherent superposition by the time of Wigner's measurement. Bob is wrong according to any interpretation which says that people \emph{can} be in superpositions - e.g. the Everett interpretation - because those interpretations say that Bob will in fact remain in a coherent superposition until Wigner's measurement if Wigner's experimental technique is sufficiently precise, and therefore it is indeed possible for Wigner's measurement to have the outcome $1$.\footnote{The Everett interpretation actually does not insist that measurements in general must have a single definite result, so in the Everett case Wigner's experiment will always have both the outcome $0$ and also the outcome $1$; but the basic structure of the argument is unchanged, as we still have Bob predicting that Wigner will not get the outcome $1$, while Wigner predicts that it's possible for him to get the outcome $1$ (and indeed, it is guaranteed that he will definitely get both outcome $0$ \emph{and} outcome $1$!). Note also that in an Everettian account of this experiment, although  one version of Bob will be wrong during the course of the measurement, his memories of making this prediction will be vanish when the two branches containing him are recombined in Wigner's measurement, so he will not subsequently be \emph{aware} of having been wrong.}

What does RQM say? In the version of RQM set out in section \ref{RQM},  it turns out that neither  Wigner nor Bob is wrong. Relative to Bob the outcome of $M$ is definitely $0$, relative to Wigner it could be either $0$ or $1$, and there is simply no more to be said about the matter. Moreover, the \textbf{internally consistent descriptions} postulate entails that if Bob gets out of the box and has a conversation with Wigner about what has come to pass, the conversation will confirm both Wigner and Bob's beliefs about the measurement outcome - and thus Bob will perceive Wigner as saying that he has obtained the outcome $0$, even if Wigner himself thinks he has said that he has obtained the outcome $1$. So not only will Bob and Wigner disagree about the outcome of a measurement, they will not even be able to rectify that disagreement by subsequently comparing notes. Wigner and Bob simply live within incommensurate realities now, and no attempt to reach across and bridge the gap can possibly succeed.  

But when postulate four is replaced with \textbf{cross-perspective links}, we  have a different story. In this approach, Bob performs his measurement and the system he measures takes on a definite value relative to him, and then perhaps he will be tempted to reason according to the steps set out in ref \cite{2018qtcc} to conclude that the measurement $M$ will definitely have outcome $0$. But relative to   Wigner, Bob is still in a superposition state and thus Wigner is able to perform a measurement on Bob in which the value of the variable measured by $M$ takes on the value $1$ relative to Wigner. Moreover, although this value is relativized to Wigner, it is also an observer-indpendent fact in the sense that other observers can find out about it by measuring or indeed just asking Wigner - for example, if Bob gets out of the box and asks Wigner what has transpired, Wigner will tell him that the measurement had outcome $1$ and Bob will perceive Wigner saying that the measurement had outcome $1$.  So Bob is wrong, although since the measurement on Bob will destroy the information stored in Bob's physical variables about the outcome of his measurement and hence also any inferences he made on the basis of that measurement, Bob will necessarily have no memory of making a wrong prediction, so no direct contradiction will ever arise. 

Note also that Bob's erroneous prediction arises only because he assumes that there is some kind of collapse when he performs his measurement. If instead he uses the version  of RQM that we have suggested in this article, then he will correctly predict that  Wigner could still get the result $1$, since he will be aware that although his observation has caused him to update his own relative state, he remains in a superposition state relative to Wigner (since that state describes the relational between Wigner and Bob, not the absolute state of Bob). So in fact no contradictory predictions will arise as long as Bob and Wigner are consistent about employing the same interpretation of quantum mechanics.

\section{Conclusion} 

In this article we have set out an updated approach to RQM including a  postulate which explicitly guarantees intersubjective agreement between observers when they perform measurements on one another.  The main motivation for our approach is to take seriously the idea that `information is physical,'  and we have argued that this principle implies that the knowledge gained by an observer when a variable becomes definite relative to them must be accessible to other observers under appropriate circumstances. We have shown that adding this postulate to RQM solves a potentially serious epistemic problem, and that it also helps answer other objections that have been made against RQM.  

Our approach also suggests some modifications to the ontology associated with RQM, because `quantum events' must now be regarded as observer-independent  in some sense, although quantum states remain relational. This suggests an ontology composed of a set of quantum events whose distribution is determined all-at-once in an external, time-symmetric way: `quantum states' are simply our best attempt at characterising the complex network of dependencies between these events, dependencies which in general will depend on the past history of interactions between an observer and system and thus also on the information that the observer possesses about the system.  Thus RQM is to be regarded as a theory of a sparse set of pointlike events or flashes, together with laws which enable us to navigate through this set of events by characterising the ways in which the joint history of a pair of systems determines the possibilities for their future interactions. 

 \section{Acknowledgements}  
 
 This publication was made possible through the support of the ID 61466 grant from the John Templeton Foundation, as part of the “The Quantum Information Structure of Spacetime (QISS)” Project (qiss.fr). The opinions expressed in this publication are those of the authors and do not necessarily reflect the views of the John Templeton Foundation.

 \bibliographystyle{unsrt}
 \bibliography{newlibrary12}{}

\end{document}